\documentclass[physrev,lengthcheck,superscriptaddress,nofootinbib,onecolumn]{revtex4-2}
\usepackage[colorlinks,allcolors=BrickRed,unicode]{hyperref}
\usepackage[dvipsnames]{xcolor}
\usepackage{amsmath,amssymb,amsfonts,physics}
\usepackage{graphicx}
\usepackage{appendix}
\pdfoutput=1

\newcommand{\eq}[1]{\begin{align}#1\end{align}}
\newcommand{\bs}{\boldsymbol}
\newcommand{\mr}{\mathrm}

\begin{document}

\title{Instantaneous normal modes of glass-forming liquids during the athermal relaxation process of the steepest descent algorithm}

\author{Masanari Shimada}

\email{masanari.shimada@torontomu.ca}

\affiliation{Department of Physics, Toronto Metropolitan University, M5B 2K3, Toronto, Canada}

\author{Kumpei Shiraishi}

\affiliation{Laboratoire Charles Coulomb (L2C), Universit\'e de Montpellier, CNRS, 34095 Montpellier, France}

\author{Hideyuki Mizuno}

\affiliation{Graduate School of Arts and Sciences, The University of Tokyo, Tokyo 153-8902, Japan}

\author{Atsushi Ikeda}

\affiliation{Graduate School of Arts and Sciences, The University of Tokyo, Tokyo 153-8902, Japan}

\affiliation{Research Center for Complex Systems Biology, Universal Biology Institute, The University of Tokyo, Tokyo 153-8902, Japan}

\date{\today}

\begin{abstract}
Understanding glass formation by quenching remains a challenge in soft condensed matter physics.
Recent numerical studies on steepest descent dynamics, which is one of the simplest models of quenching, revealed that quenched liquids undergo slow relaxation with a power law towards mechanical equilibrium and that the late stage of this process is governed by local rearrangements of particles.
These advances motivate the detailed study of instantaneous normal modes during the relaxation process because the glassy dynamics is considered to be governed by stationary points of the potential energy landscape.
Here, we performed a normal mode analysis of configurations during the steepest descent dynamics and found that the dynamics is driven by almost flat directions of the potential energy landscape at long times.
These directions correspond to localized modes and we characterized them in terms of their statistics and structure using methods developed in the study of local minima of the potential energy landscape.
\end{abstract}

\maketitle 

\section{Introduction}

Glasses are typically made by quenching liquids~\cite{Cavagna2009Supercooled,Berthier2011Theoretical}.
The non-equilibrium relaxation dynamics after quenching\footnote{
In this paper, we use the word ``quench'' to refer to the operation of instantaneously lowering the temperature and ``relaxation'' to the non-equilibrium dynamics after quenching.} is highly non-trivial and has attracted much interest in the field of glassy systems.
The equilibration time of quenched liquids exceeds the experimental timescale below the glass transition temperature $T_g$, i.e., quenched liquids are always out of equilibrium. 
The complexity of this process is most vividly demonstrated in a phenomenon called aging~\cite{Cugliandolo1993Analytical,Kurchan1996Phase,Kob2000Aging}.
Below $T_g$, the two-time correlation functions depend on the waiting time after the system is quenched.
This behavior is usually not observed in equilibrium systems and is a characteristic of non-equilibrium glassy systems.

To further understand the relaxation after quenching, recent numerical studies have focused on steepest descent dynamics~\cite{Chacko2019Slow,Nishikawa2021Relaxation,GonzalezLopez2020energy,Folena2021Gradient}.
This is the Langevin equation with the noise term suppressed and one of the simplest models of quenching to zero temperature.
This algorithm is of particular importance because there is, at least in principle, a one-to-one correspondence between initial states and mechanically stable final states, called inherent structures.
In Ref.~\cite{Chacko2019Slow}, packings of athermal particles slightly above jamming were studied in both two and three spatial dimensions.
The authors of this study focused on the root mean squared particle speed, which is a square root of the mean squared force 
\eq{
    W = \frac{1}{N} \sum_{i=1}^N \bs{f}_i^2 ,
}
where $N$ is the number of particles and $\bs{f}_i$ is the force acting on particle $i$.
This $W$-function was first introduced to study the so-called geometrical transition discussed in the context of the mode-coupling theory~\cite{Angelani2000Saddles,Broderix2000Energy}.
It was found in Ref.~\cite{Chacko2019Slow} that the root mean squared particle speed decays with a power law $\propto t^{-\beta}$, where $t$ is the time after quenching.
The exponent $\beta$ depends on the spatial dimension: $\beta=0.92$ for two dimensions and $\beta=0.85$ for three dimensions.
This power-law relaxation of liquids into inherent structures is governed by temporally intermittent and spatially localized spots of non-affine displacements, which are called hot spots in Ref.~\cite{Chacko2019Slow}.
The hot spots are surrounded by backgrounds exhibiting swirling motion and become more prominent and sparser with time, resembling local plastic events in glasses under shear known as shear transformation zones~\cite{Argon1979Plastic,Falk1998Dynamics,Tanguy2006Plastic,Schall2007Structural}.

Soon after the research reported in Ref.~\cite{Chacko2019Slow}, Nishikawa \textit{et al.} investigated the steepest descent dynamics in several glass-forming models over a broad range of temperatures\footnote{
There are two temperatures to characterize quenching: the parent and target temperatures.
The former is the one at which the system is equilibrated before quenching and the latter is the one to which the system is quenched.
In the case of the steepest descent dynamics, the latter is zero and thus the word ``temperature'' always means the parent temperature in this paper.
} and spatial dimensions~\cite{Nishikawa2021Relaxation}. 
The results showed that a mean-field Mari-Kurchan (MK) model undergoes a qualitative change in the relaxation dynamics. 
The root mean squared particle speed follows a power-law decay at high temperatures and an exponential decay at low temperatures.
The exponent of the power-law decay at high temperatures is $\beta = 0.75$, which is different from the ones mentioned in the previous paragraph~\cite{Chacko2019Slow}.

In contrast, the root mean squared particle speed decays algebraically even much below the mode-coupling transition temperature $T_{\mr{MCT}}$ in all the finite-dimensional models considered in Ref.~\cite{Nishikawa2021Relaxation}.
The exponent of this decay is the same as $\beta$ reported in Ref.~\cite{Chacko2019Slow} at high temperatures, $\beta=0.92$ for two dimensions and $\beta=0.85$ for three dimensions.
This exponent varies with the temperature and spatial dimension, but it is universal for systems with different pairwise interactions.
It was found that these relaxation processes of the finite-dimensional models are dominated by spatially localized hot spots\footnote{
In Ref.~\cite{Nishikawa2021Relaxation}, hot spots are called defects using a different but equivalent definition.
} as observed in Ref.~\cite{Chacko2019Slow}.
The density of the hot spots rapidly decreases with decreasing temperature, but it remains finite and does not exhibit singular behavior. 
This means that hot spots dominate the relaxation dynamics at all temperatures in the finite-dimensional models.

The time evolution of the mean squared particle speed was also investigated theoretically using a mean-field spin glass model, the mixed $p$-spin model~\cite{Folena2020Rethinking}.
The transition from the power-law to exponential decay observed in the MK model~\cite{Nishikawa2021Relaxation} was predicted though the power-law exponent $\beta = 0.83$ is slightly different from the one of the MK model.
The transition temperature is called the state-following temperature $T_{\mr{SF}}$ and the exponent $\beta$ does not depend on the temperature above $T_{\mr{SF}}$.
It was also pointed out that a temperature called the onset temperature $T_{\mr{onset}} ( > T_{\mr{MCT}})$ separates a high-temperature memoryless regime and a low-temperature memorious regime.

In view of this progress, it is essential to explore the spatial structure of the hot spots in depth.
Unlike equilibrium liquids, it is predicted that configurations during relaxation are close to stationary points of the potential energy landscape~\cite{Kurchan1996Phase}.
At long times, almost all directions of such stationary points are stable; namely, their curvatures are positive.
The dynamics of the system is driven by few almost flat directions with positive and negative curvatures and the relaxation proceeds slowly along a ``gorge'' of the potential energy landscape.
Thus, investigating instantaneous normal modes, whose eigenvalues give the curvatures of the directions of the corresponding eigenvectors, of these transient stationary points is a reasonable approach to study hot spots.
These regions with high mobility will be linked to almost flat but slightly unstable directions.
Although several studies have investigated instantaneous normal modes of equilibrium liquids~\cite{Seeley1989Normal,Broderix2000Energy,Ciliberti2004Localization,Grigera2006Geometrical,Cavagna2009Supercooled,WidmerCooper2009Localized,Clapa2012Localization,Palyulin2018Parameter} and glasses under finite-rate shear~\cite{Oyama2021Instantaneous}, there are relatively few studies~\cite{Donati2000Role} that explored instantaneous modes during the relaxation process after quenching.

Contrary to instantaneous normal modes, there have been a large number of studies on normal modes of inherent structures.
Since the low-frequency modes of inherent structures have been extensively studied~\cite{Lerner2016Statistics,Mizuno2017Continuum,Lerner2021Low}, it is important to compare them with instantaneous modes during the athermal relaxation described above.
In the lowest-frequency tail of the spectrum of inherent structures, spatially localized modes whose cores are surrounded by algebraically decaying far fields are observed for a broad class of glasses, in addition to acoustic phonons~\cite{Lerner2016Statistics}.
These modes, known as quasi-localized modes, are attributed to several anomalous low-energy properties of glasses, including plastic events under shear~\cite{Maloney2006Amorphous,Tanguy2010Vibrational,Manning2011Vibrational} and low-temperature thermal properties~\cite{Zeller1971Thermal,Anderson1972Anomalous}.
Recently, comparing the quasi-localized modes and unstable modes found at saddle points, it was also established that these two types of modes are deeply interlinked in terms of their structures~\cite{Shimada2021Spatial}.
Therefore, it is meaningful to analyze instantaneous modes during relaxation using the methods adopted to analyze the quasi-localized modes.

In this study, we explored instantaneous normal modes of transient configurations during relaxation using typical glass-forming liquids.
Instead of the ensemble average at a fixed time, we considered the average at a fixed mean squared force $W$.
Since many observables exhibit rapid time dependences in the final stage of the dynamics, the latter average is easier to treat.
Using numerical methods that have been developed in the study of inherent structures, we investigated unstable instantaneous normal modes, which can be seen as ``bare'' hot spots~\cite{Chacko2019Slow,Nishikawa2021Relaxation}.
We verified that the dynamics is driven by few almost flat directions of the potential energy landscape as predicted in Ref.~\cite{Kurchan1996Phase} and found that these directions correspond to unstable localized modes, particularly in the late stage in accordance with the previous observations~\cite{Chacko2019Slow,Nishikawa2021Relaxation}.
Furthermore, focusing on the most unstable mode of each sample, we revealed that its core remains mechanically unstable while the overall system, i.e., the core plus far field, is rapidly stabilized.
The correlation length of the far field follows a non-trivial stretched exponential law.
These findings can help us understand the steepest descent dynamics by bridging two different descriptions: one based on real space and the other based on the potential energy landscape.

This paper is organized as follows.
In Sec.~\ref{sec:model}, we delineate the model and numerical methods.
In Sec.~\ref{sec:mean}, we investigate the time evolution of the mean squared force and the fraction of unstable modes.
In Sec.~\ref{sec:spectrum}, we investigate the unstable part of the spectrum and perform a finite-size scaling analysis to distinguish localized and delocalized modes.
In Sec.~\ref{sec:most}, we focus on the most unstable mode and analyze its spatial structure.
In Sec.~\ref{sec:conclusion}, we conclude this paper.
In Appendix~\ref{sec:system}, we examine the system size dependence of several quantities.
In Appendix~\ref{sec:raw}, we show raw data for the participation ratio, which is used for the finite-size scaling analysis in Sec.~\ref{sec:spectrum}.
In Appendix~\ref{sec:definitions}, we give the detailed definitions of the decay and energy profiles introduced in Sec.~\ref{sec:most}.
In Appendix~\ref{sec:number}, we show the data that suggest that the most unstable mode of a configuration has only a single core.
In Appendix~\ref{sec:jammed}, we investigate weakly jammed packings interacting via a harmonic potential and show that the results in the main text do not change qualitatively.

\section{Model and methods}\label{sec:model}

\begin{figure}
    \centering
    \includegraphics{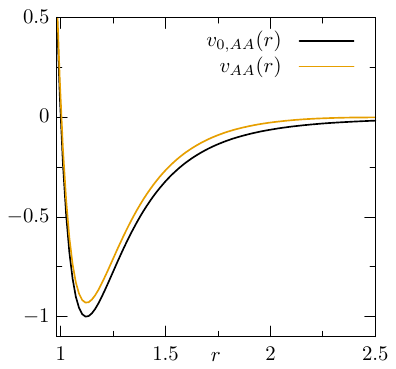}
    \caption{
    The original KA potential in Eq.~\eqref{KALJ} and the modified potential in Eq.~\eqref{smoothed KALJ} between species A and A.
    }
    \label{fig:potentials}
\end{figure}

Our model is based on the three-dimensional Kob-Andersen binary Lennard-Jones (KALJ) model~\cite{Kob1995Testing,Kob1995Testinga}, which has been widely used in the context of the glass transition because of its particularly high glass-forming ability.
This model has two types of particles, $A$ and $B$, with a composition ratio of 80:20.
Our system is a cubic box with periodic boundary conditions applied in all directions.
Both types of particles have the same mass $m$, and the particles interact with each other via a pair potential
\eq{
    v_{0,ab} (r) = 4 \epsilon_{ab} \left[ \left( \frac{\sigma_{ab}}{r} \right)^{12} - \left( \frac{\sigma_{ab}}{r} \right)^6 \right] , \label{KALJ}
}
where $r$ is the distance between a pair and the parameters are as follows: $\epsilon_{AB}=1.5\epsilon_{AA}$, $\epsilon_{BB}=0.5\epsilon_{AA}$, $\sigma_{AB}=0.8\sigma_{AA}$, and $\sigma_{BB}=0.88\sigma_{AA}$.
In this paper, length, mass, and time are reported in units of $\sigma_{AA}$, $m$, and $\sqrt{m\sigma_{AA}^2/\epsilon_{AA}}$, respectively.
The total number density is fixed at $\rho = 1000/(9.4)^3 \approx 1.204$~\cite{Kob1995Testing,Kob1995Testinga}.
In the KALJ model, Eq.~\eqref{KALJ} is conventionally cut at a distance $r_{\mr{cut},ab} = 2.5\sigma_{ab}$.
However, the discontinuity of the pair force $v_{0,ab}'$ at the cutoff distance strongly affects the low-frequency modes~\cite{Shimada2018Anomalous}. 
Thus, we modified Eq.~\eqref{KALJ} so that its derivative tends to zero continuously 
\eq{
    v_{ab} (r) = 
    \begin{cases}
        v_{0,ab} (r) - v_{0,ab} (r_{\mr{cut},ab}) - ( r - r_{\mr{cut},ab} ) v'_{0,ab} (r_{\mr{cut},ab}) & ( r < r_{\mr{cut},ab} ), \\
        0 & ( r > r_{\mr{cut},ab} )
    \end{cases}
    . \label{smoothed KALJ}
}
Compare $v_{0,AA}(r)$ and $v_{AA}(r)$ in Fig.~\ref{fig:potentials}.
Using Eq.~\eqref{smoothed KALJ}, we carried out molecular dynamics (MD) simulations.
First, the system was equilibrated at a sufficiently high temperature $T=2.0$ in the NVT ensemble for a time interval of $t=500$, and then we performed simulations in the NVE ensemble\footnote{
The choice of the ensemble is merely a matter of taste.
Even if we use other ensembles, the results will not change as long as the initial states are equilibrated.
We tried to make the procedure as simple as possible and thus performed NVE simulations finally.
} for the same time interval.
At this high temperature, the equilibration time is about unity.
Independent liquid configurations were sampled in the subsequent NVE simulation.
The time interval between two consecutive configurations is $t=50$.
Note that in Sec.~\ref{sec:low temperature} we will show the results at much lower temperatures to compare them with the results in Sec.~\ref{sec:high temperature}.
The algorithm for generating these low-temperature samples is explained in detail at the beginning of Sec.~\ref{sec:low temperature}.

\begin{table}
    \renewcommand{\arraystretch}{1.2}
    \setlength\tabcolsep{0.5em}
    \centering
    \caption{
    The number of samples for each $N$.
    The number of samples used to plot Figs.~\ref{fig:coredist} and \ref{fig:corecumulative} is given in parentheses.
    See the text for details.
    }
    \begin{tabular}{cc} \hline \hline
        $N$     & \#samples     \\ \hline
        1000    & 1000 (10000)  \\
        2000    & 1000          \\
        4000    & 1000          \\
        8000    & 500           \\
        32000   & 498 (10000)   \\
        128000  & 495           \\ \hline \hline
    \end{tabular}
    \label{tab:samples}
\end{table}

Starting from these liquid states, we performed simulations of the steepest descent dynamics~\cite{Chacko2019Slow,Nishikawa2021Relaxation}; namely, we numerically solved the following equation of motion:
\eq{
    \gamma \dv{\bs{r}_i}{t} = - \pdv{V}{\bs{r}_i} , \label{steepest descent}
}
where $\bs{r}_i$ is the position of particle $i$, and $V = \sum_{i<j} v_{ab}(r_{ij})$ is the total potential energy.
The discretized time step was set at $\Delta t = 0.0001$ and the friction coefficient $\gamma$ at unity.
This dynamics stops at the time when the system reaches an inherent structure, at which $\pdv*{V}{\bs{r}_i} = 0$ for all $i$, but since our purpose was to investigate transient configurations during the dynamics, we did not need to wait for the system to reach an inherent structure.
The number of configurations used for each $N$ is listed in Tab.~\ref{tab:samples}.
To plot Figs~\ref{fig:coredist} and \ref{fig:corecumulative}, we needed a particularly large sample size and the number of configurations is given in the parentheses in Tab.~\ref{tab:samples}.
All simulations were performed using Large-scale Atomic/Molecular Massively Parallel Simulator (LAMMPS)~\cite{Plimpton1995Fast,Thompson2022LAMMPS}.

We performed a normal mode analysis~\cite{Kittel2004Introduction} of instantaneous configurations during the steepest descent dynamics.
The dynamical matrix, the Hessian matrix of the total potential $V$, was diagonalized to obtain its eigenvalues $\lambda_\alpha$ and eigenvectors $\bs{e}_\alpha = (\bs{e}_{\alpha,1},\dots,\bs{e}_{\alpha,N})$, where $\alpha=1,2,\dots,3N-3$.
The three modes corresponding to the global translations were excluded.
The eigenvalues are sorted in ascending order: $\lambda_1<\lambda_2<\dots<\lambda_{3N-3}$, and the eigenvectors are normalized as $|\bs{e}_{\alpha}|=1$.

\section{Results}\label{sec:results}

\subsection{Mean squared force and fraction of unstable modes}\label{sec:mean}

In this subsection we investigate the time evolution of the mean squared force $W$ and the fraction of unstable modes
\eq{
    f_u = \int_{-\infty}^0 \dd{\lambda} D(\lambda),
}
where $D(\lambda)$ is the spectrum, or the density of states, defined as 
\eq{
    D (\lambda) = \frac{1}{3N-3} \sum_{\alpha=1}^{3N-3} \delta \qty ( \lambda - \lambda_\alpha ) . \label{spectrum}
}
As mentioned in the introduction, the mean squared force $W$ is an important quantity to characterize the steepest-descent dynamics and we will first confirm that our results are consistent with the previous results~\cite{Chacko2019Slow,Nishikawa2021Relaxation} below by using $W$.

However, we emphasize that our main purpose in this subsection is \emph{not} to determine the detailed time-dependence of $W$, which has been studied intensively in Refs.~\cite{Chacko2019Slow,Nishikawa2021Relaxation}, but to investigate the qualitative relation between $W$ and $f_u$.
In particular, we verify the prediction in Ref.~\cite{Kurchan1996Phase}; namely, the relaxation dynamics after quenching proceeds slowly because the system travels along an almost flat gorge in the potential energy landscape.
From $W$ and $f_u$, one can estimate the dependence of the potential energy $V$ on $f_u$ at long times~\cite{Cavagna2009Supercooled,Broderix2000Energy,Grigera2002Geometric,Grigera2006Geometrical}, which is necessary to evaluate the flatness of the potential energy landscape as will be described in the next paragraph.

The $f_u$-derivative of the potential energy $\pdv*{(V/N)}{f_u}$ gives the difference of the potential energy between stationary points with different numbers of unstable modes and has been well studied in terms of saddle modes~\cite{Cavagna2001Fragile,Cavagna2009Supercooled,Broderix2000Energy,Grigera2002Geometric,Grigera2006Geometrical}.
We have
\eq{
    V_{n+1} - V_n & = \frac{1}{3N-3} \frac{V_{n+1} - V_n}{f_{u,n+1} - f_{u,n}} = \frac{N}{3N-3} \frac{V_{n+1}/N - V_n/N}{f_{u,n+1} - f_{u,n}} \xrightarrow{N\to\infty} \frac{1}{3} \pdv{(V/N)}{f_u} , \label{dvdfu}
}
where $\bullet_{n}$ denotes the value measured at a stationary point with $n$ unstable directions, i.e., modes with negative eigenvalues.
The first equality is a consequence of $f_{u,n} = n/(3N-3)$.
The derivative in Eq.~\eqref{dvdfu} measures how much energy the system releases when an unstable mode disappears and the system is slightly stabilized.
This is the meaning of the flatness, i.e., the smaller this derivative, the flatter the potential energy landscape.
Note that, for the rightmost expression in Eq.~\eqref{dvdfu} to be meaningful, we assumed $V_{n+1}-V_n = \order{N^0}$, which is plausible for finite-dimensional systems because the dynamics is driven by localized modes as will be shown in Sec.~\ref{sec:spectrum}.
Also, Eq.~\eqref{dvdfu} is strictly justified only for stationary points and should be used as an approximation for instantaneous normal modes.
We nevertheless expect that this approximation improves as $t\to\infty$ during the steepest descent dynamics because the system stays close to stationary points for a long time in the last stage of the dynamics~\cite{Kurchan1996Phase}.
Thus in this section, we investigate the flatness of the potential energy landscape at long times using Eq.~\eqref{dvdfu}.

In Sec.~\ref{sec:high temperature}, we explain the difficulty in computing the ensemble average at a fixed time and introduce an ensemble with a fixed $W$.
Then we show high-temperature results at $T=2.0$ and discuss $\pdv*{(V/N)}{f_u}$ in the limit $t\to\infty$.
In Sec.~\ref{sec:low temperature}, we show the same results at low temperatures.

\subsubsection{High temperature}\label{sec:high temperature}

\begin{figure}
    \centering
    \includegraphics{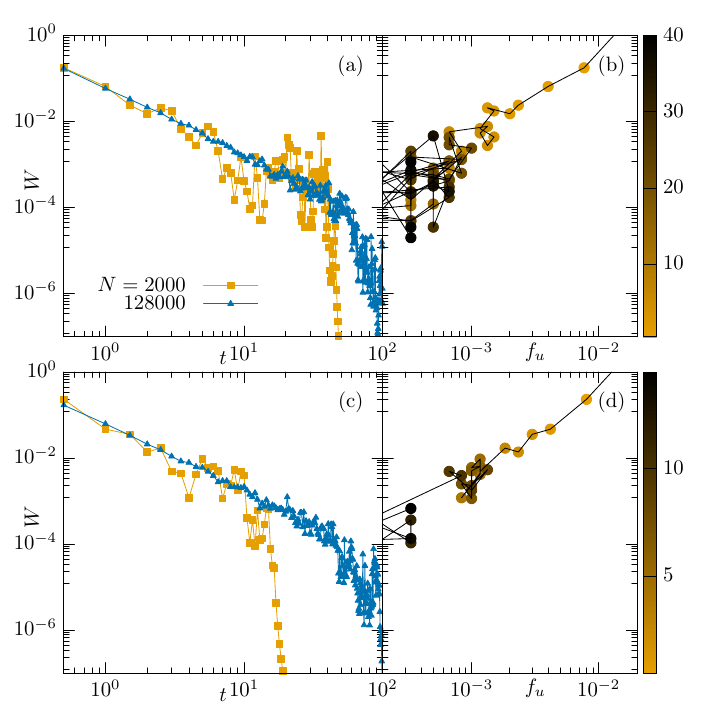}
    \caption{
    (a, c) Mean squared force $W$ for $N=2000$ and 128000 as functions of the time.
    (b, d) Mean squared force $W$ versus fraction $f_u$ of unstable modes corresponding to the data for $N=2000$ in the panels (a, c).
    The color bars indicate the time.
    }
    \label{fig:raw}
\end{figure}

We first show raw data at $T=2.0$ in Fig.~\ref{fig:raw} to introduce the averaging procedure adopted in this study.
Figure~\ref{fig:raw}(a) shows the mean squared force $W$ for $N=2000$ and 128000 as functions of the time and Fig.~\ref{fig:raw}(b) is a parametric plot of $W$ versus the fraction of unstable modes $f_u$.
The data in Fig.~\ref{fig:raw}(b) were taken from the same trajectory of $N=2000$ in Fig.~\ref{fig:raw}(a).
In Fig.~\ref{fig:raw}(c, d), we present the same plots for different samples, to be compared in the next paragraph.
Although the data in Fig.~\ref{fig:raw} is noisy, one can see that the mean squared force $W$ first exhibits an almost algebraic decay followed by a precipitous drop at a particular time point.
The system finds an inherent structure and the dynamics stops immediately after the drop.
For larger $N$, the range of the algebraic decay extends and the convergence to an inherent structure slows down.
The fraction of unstable modes decreases with decreasing $W$, but the decay is not monotonic and the value of $f_u$ oscillates randomly.

Figure~\ref{fig:raw} indicates the difficulty in carrying out an ensemble average; because of the strong fluctuations of the time at which $W$ converges to zero, the ensemble average at a fixed time will suffer from significant uncertainties in the late stage.
Compare Fig.~\ref{fig:raw}(a,b) and (c,d).
In addition, we point out a conceptual problem of the fixed time average.
Because of the fluctuations of the convergence time, there is a time interval, in which some samples already converged to inherent structures while the others have not converged yet and are migrating in the potential energy landscape. 
If we adopt the fixed time average, these two kinds of samples are mixed within this interval.
However, they are in physically different states and such mixing may lead to unclear, or possibly incorrect, results because instantaneous normal modes strongly reflect the structure of the potential energy landscape.
To avoid these issues, we averaged the data at fixed $W$.
During simulations, we sampled a configuration every time $W$ reaches a specific value and averaged observables over configurations with the same $W$.

\begin{figure}
    \centering
    \includegraphics{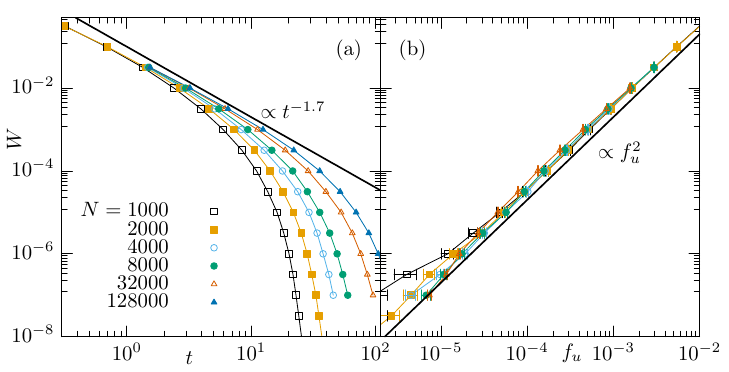}
    \caption{
    (a) Average mean squared force as functions of the time.
    The solid line represents a power law $W \propto t^{-1.7}$~\cite{Chacko2019Slow,Nishikawa2021Relaxation}.
    Note that $W$ is controlled in this plot, see the text for details.
    The errors are smaller than the symbols and thus not shown.
    (b) Average mean squared force versus fraction of unstable modes.
    The solid line represents a power law $W \propto f_u^2$.
    }
    \label{fig:force}
\end{figure}

Figure~\ref{fig:force} shows the same data as in Fig.~\ref{fig:raw}, which are averaged with $W$ fixed.
The average mean squared force in Fig.~\ref{fig:force}(a) decays algebraically and drops mildly at a time depending on the system size $N$.
The algebraic decay in the middle stage is compared by a power law $W \propto t^{-1.7}$~\cite{Chacko2019Slow, Nishikawa2021Relaxation}, as shown by the solid line.
Note that this is twice the exponent $\beta=0.85$ reported in Refs.~\cite{Chacko2019Slow,Nishikawa2021Relaxation} because $\sqrt{W}$ was measured in these studies.
Since it is difficult to investigate how the power-law relation $W\sim t^{-1.7}$ breaks down in the late stage of the dynamics for our limited system sizes, understanding this final regime quantitatively is left for future work.
Figure~\ref{fig:force}(b) shows another power law, $W \propto f_u^2$, which holds down to $W \sim 10^{-8}$.
The system approaches a stable inherent structure $W\to0$ as the fraction of unstable modes $f_u$ becomes smaller.
It seems that the power law $W \propto f_u^2$ breaks down slightly in the smallest-$W$ region $W \sim 10^{-8}$.
One needs much larger systems to establish the precise functional form $W(f_u)$ in this region.
However, the details of these data will not affect the analysis of the $f_u$-derivative of the potential energy described at the beginning of this section.
Likewise, the exponents of the power laws observed in Fig.~\ref{fig:force} are not generally universal as will be discussed in Sec.~\ref{sec:low temperature} (see also Refs.~\cite{Chacko2019Slow,Nishikawa2021Relaxation}), but the results below do not strongly depend on the values of the exponents.

Related to these results, we note that similar scaling relations were reported for the spherical Sherrington-Kirkpatrick (SK) model~\cite{Kurchan1996Phase}.
This simple mean-field model does not show a glass transition, but it is characterized by slow relaxation similar to that of the spherical $p$-spin model with $p>2$, which has a spin-glass phase.
In Ref.~\cite{Kurchan1996Phase}, it was shown that $W\sim t^{-2}$ and $f_u\sim t^{-3/2}$, which leads to $W\sim f_u^{4/3}$ in the spherical SK model.

We apply Eq.~\eqref{dvdfu} to the regime after the power law $W\sim t^{-1.7}$ in Fig.~\ref{fig:force}(a).
From Eq.~\eqref{steepest descent}, we have
\eq{
   \gamma \qty | \dv{(V/N)}{t} | = -\frac{\gamma}{N} \sum_{i=1}^N \pdv{V}{\bs{r}_i} \dv{\bs{r}_i}{t} = \frac{1}{N} \sum_{i=1}^N \bs{f}_i^2 = W .
}
Using the power law\footnote{As pointed out above, this power law might break down slightly in the smallest-$W$ regime and its exponent is model and temperature dependent (see Sec.~\ref{sec:low temperature} and Appendix~\ref{sec:jammed}), but this does not affect the discussion in this paragraph unless the value of the limit $\lim_{t\to\infty}\pdv*{(V/N)}{f_u}$ changes.} $W\propto f_u^2$, we obtain
\eq{
    \pdv{(V/N)}{f_u} = \frac{\dv{(V/N)}{t}}{\dv{f_u}{t}} \sim - \frac{W}{\frac{1}{\sqrt{W}}\dv{W}{t}} = \frac{W^{3/2}}{\qty|\dv{W}{t}|} .
}
As shown in Fig.~\ref{fig:force}(a), the mean squared force finally decays to zero very fast.
For example, one can assume an exponential decay $W(t) \sim e^{-\tau/\tau_0}$ or an algebraic decay $W(t) \sim t^{-\upsilon}$ with a sufficiently large exponent $\upsilon>2$.
Thus, it is reasonable to assume that $W^{3/2}\ll\qty|\dv*{W}{t}|$, which leads to $\pdv*{(V/N)}{f_u}\xrightarrow{t\to\infty} 0$.
The fact that the $f_u$-derivative of the potential energy tends to zero in the final stage of the dynamics indicates that the potential energy landscape becomes flatter as the system approaches an inherent structure~\cite{Cavagna2009Supercooled} as explained above.
This is direct evidence of the almost flat gorge of the potential energy landscape predicted in Ref.~\cite{Kurchan1996Phase}.
We again emphasize that the details of the functional forms of $W(t)$ and $W(f_u)$ in Fig.~\ref{fig:force} are not relevant unless the value of the limit $\lim_{t\to\infty}\pdv*{(V/N)}{f_u}$ changes.

To conclude this subsection, we make a remark on the measurement of the potential energy $V$.
As seen in the analysis of $\pdv*{(V/N)}{f_u}$, the most important observable is the potential energy itself rather than the mean squared force $W$.
However, the potential energy does not vanish at inherent structures in general and its value $V_{\mr{inh}}$ fluctuates while $W$ converges to zero.
The measurement of $W$ can be more precise than that of $V$.
This is a reason why the $W$-function was chosen in this and related studies~\cite{Chacko2019Slow,Nishikawa2021Relaxation}.
Also, this $W$-function has been used to locate saddle points in the potential energy landscape to verify the prediction of the mode-coupling theory as mentioned in the introduction~\cite{Angelani2000Saddles,Coslovich2019localization}.

\subsubsection{Low temperatures}\label{sec:low temperature}

Here, we present the results at low temperatures $T=0.4500$, 0.4308, and 0.3971, which correspond to those in Fig.~\ref{fig:force}, to investigate how the above results change at sufficiently low temperatures\footnote{
In this subsection, we set the total number density to $\rho=1.204$. Thus, the linear size of the system is $L\approx9.3999$, which is slightly different from the original value $L=9.4$ in Refs.~\cite{Kob1995Testing,Kob1995Testinga}.
}.
For $T=0.4500$, equilibrium configurations were prepared by the MD simulations in the NVT ensemble using the Nos\'e-Hoover thermostat~\cite{Frenkel2001Understanding}.
Starting from configurations equilibrated at $T=2.0$, the MD simulations in the NVT ensemble were performed for $t = 10^5$, and then we performed simulations in the NVE ensemble for the same time interval.
We used time-reversible integrator~\cite{Martyna_1996,Frenkel2001Understanding} with a time step of 0.005.
The number of samples prepared at this temperature is 1000.
To sample the equilibrium configurations at lower temperatures, $T=0.4308$ and 0.3971, we used the replica exchange method~\cite{Hukushima1996Exchange,Yamamoto2000Replica}.
In our MD simulations, 18 replicas were used and each replica corresponded to a temperature ranging from 0.7800 to 0.3971.
Exchange trials were performed every 2800 MD steps using the Metropolis criterion.
We checked that all replicas were equilibrated after 350000 exchange trials and sampling was performed every 10000 trials thereafter.
All calculations were carried out in an in-house code that uses MPI to handle parallel computations of exchanging replicas.
Because of the high computational cost, we were only able to prepare 45 samples of $N = 1000$.

\begin{figure}
    \centering
    \includegraphics{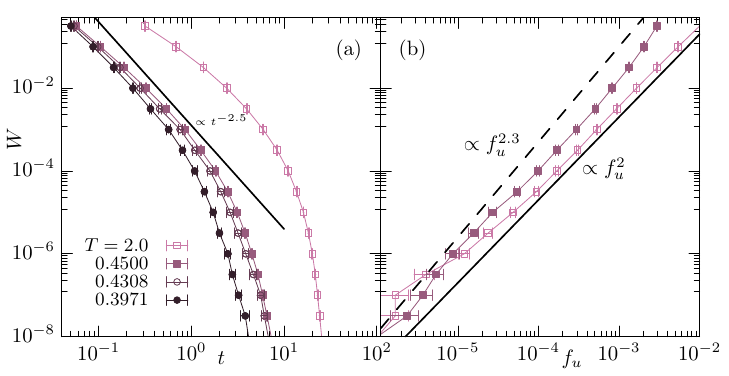}
    \caption{
    (a) Time evolution of mean squared force from different temperatures $T$.
    The system size is $N=1000$.
    The solid line represents a power law $W \propto t^{-2.5}$.
    (b) Mean squared force versus fraction of unstable modes.
    The solid and dashed lines represent power law $W \propto f_u^2$ and $W \propto f_u^{2.3}$, respectively.
    }
    \label{fig:remd_force}
\end{figure}

Using these low-temperature configurations, we performed the same analysis as described in the previous subsection.
Figure~\ref{fig:remd_force} corresponds to Fig.~\ref{fig:force}.
The solid line in Fig.~\ref{fig:remd_force}(a) represents a power law $W\propto t^{-2.5}$, which was observed in Ref.~\cite{Nishikawa2021Relaxation}.
This is not inconsistent with the data, though we cannot verify this power law because of the small system size.
Also, it is possible to generalize the discussion of $\pdv*{(V/N)}{f_u}$.
Setting $W\sim f_u^b$, where $b>0$, yields
\eq{
    \pdv{(V/N)}{f_u} \sim \frac{W^{2-1/b}}{\qty|\dv{W}{t}|} .
}
As shown in Fig.~\ref{fig:remd_force}(a), $W$ drops similarly to the high-temperature data in Fig.~\ref{fig:force}(a) in the last stage of the dynamics and thus we conclude that $\pdv*{(V/N)}{f_u}\xrightarrow{t\to\infty} 0$.
Larger systems are necessary to establish this result.

In summary, the data in Figs.~\ref{fig:force} and \ref{fig:remd_force} indicate that the system moves on an almost flat gorge of the potential energy landscape at long times, which is consistent with the results of mean-field models in Ref.~\cite{Kurchan1996Phase}.
This result does not strongly depend on the precise functional forms of $W(t)$ and $W(f_u)$, which may not be universal.
In Appendix~\ref{sec:jammed}, we will show the corresponding results for weakly jammed solids and confirm that $\lim_{t\to\infty}\pdv*{(V/N)}{f_u} = 0$, though the specific exponents of the power laws change.

\subsection{Spectrum and localization}\label{sec:spectrum}

\begin{figure}
    \centering
    \includegraphics{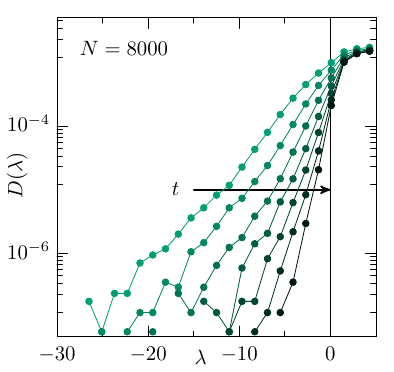}
    \caption{
    Unstable parts of the spectra for $N=8000$ during the dynamics.
    The values of the mean squared force are $\log_{10}W = -4.5,-4.0,-3.5,-3.0,-2.5,-2.0$, and $-1.5$ from right to left.
    The arrow indicates the direction of time, which is the opposite of $W$.
    } 
    \label{fig:dos_8k}
\end{figure}

The spectrum $D(\lambda)$ in Eq.~\eqref{spectrum} is a fundamental quantity for characterizing stationary points of the potential energy landscape.
Though Figs.~\ref{fig:raw} and \ref{fig:force} show an integrated version of it, $f_u$, direct investigation of the spectrum gives more insight into the dynamics.
In this subsection, we also investigate the degree of localization of unstable modes and show that all unstable modes are localized at long times, which is consistent with the fact that the relaxation dynamics is governed by spatially localized hot spots~\cite{Chacko2019Slow,Nishikawa2021Relaxation}.

Figure~\ref{fig:dos_8k} shows unstable parts of the spectra for $N=8000$ during the dynamics.
As in Sec.~\ref{sec:mean}, each spectrum is labeled by the value of $W$ and the specific values are $\log_{10}W = -4.5,-4.0,-3.5,-3.0,-2.5,-2.0$, and $-1.5$ from right to left.
Note that, however, the arrow in Fig.~\ref{fig:dos_8k} indicates the direction of time, which is the opposite of $W$.
In Appendix~\ref{sec:system}, we will confirm that the spectrum does not depend on the system size $N$.
From Fig.~\ref{fig:dos_8k}, one can see that the unstable part of the spectrum always shows an exponential tail.

\begin{figure}
    \centering
    \includegraphics{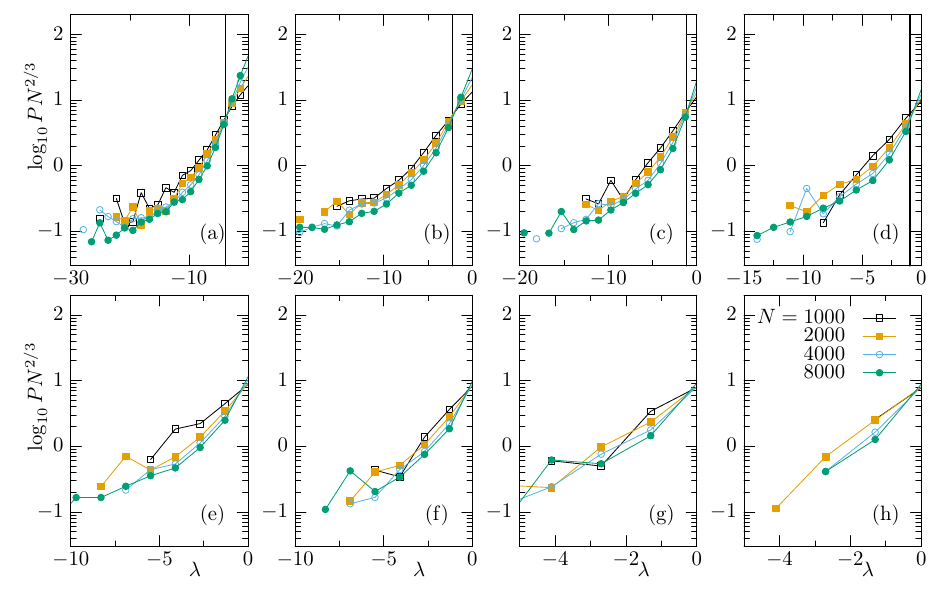}
    \caption{
    $\log_{10}PN^{2/3}$ versus $\lambda$.
    The values of $\log_{10}W$ are (a) $-1.5$, (b) $-2$, (c) $-2.5$, (d) $-3$, (e) $-3.5$, (f) $-4$, (g) $-4.5$, and (h) $-5$.
    For panels (a), (b), (c), and (d), the estimated mobility edges are shown by the solid lines.
    }
    \label{fig:edge}
\end{figure}

\begin{figure}
    \centering
    \includegraphics{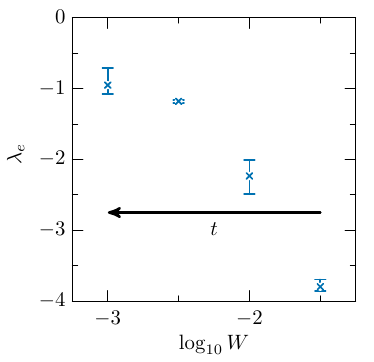}
    \caption{
    Mobility edge as a function of $W$.
    The arrow indicates the direction of time, which is the opposite of $W$.
    The errors are estimated as minimum and maximum values of the intersections of pairs of data for different $N$ in Fig.~\ref{fig:edge}.
    }
    \label{fig:edge_W}
\end{figure}

The exponential tail of the spectrum is a characteristic of localized modes found at saddle points~\cite{Coslovich2019localization}.
The level spacing distribution of such localized modes is a Poisson distribution, which means that these eigenvalues are uncorrelated with each other~\cite{Coslovich2019localization}.
These saddle modes are truly localized, meaning that their particle motions strongly decay far from the localized cores, while the quasi-localized modes found at inherent structures have algebraically decaying far fields and its density follows a quartic law $g(\omega^4) = \omega D(\omega^2) \propto \omega^4$, where $\omega=\sqrt{\lambda}$ is the frequency~\cite{Lerner2016Statistics,Mizuno2017Continuum}.
Thus it is important to understand the degree of localization of the unstable modes observed in Fig.~\ref{fig:dos_8k}.
This motivated us to perform a finite-size scaling analysis of the participation ratio~\cite{Mazzacurati1996Low,Schober2004Size,Taraskin1999Anharmonicity}
\eq{
    P_\alpha = \frac{1}{N} \left( \sum_{i=1}^N |\bs{e}_{\alpha,i}|^4 \right)^{-1} ,
}
which measures the degree of localization of a given mode $\alpha$.
If all particles vibrate equally, $P_\alpha=1$ and if only a single particle vibrates, $P_\alpha=1/N$.
Thus, $NP_\alpha$ gives the number of particles participating in the mode,
enabling us to determine the mobility edge $\lambda_e$, which separates localized and delocalized modes.
Specifically, the average participation ratio $NP= N\langle P_\alpha\rangle_\lambda$ at $\lambda$, for delocalized modes grows at least linearly with $L$, where $L$ is the linear size of the system, whereas it is independent of $L$ for localized modes~\cite{Shklovskii1993Statistics,Coslovich2019localization}.
Namely, delocalized modes expand throughout the system at least in a direction\footnote{
This interpretation should not be taken too seriously.
Since we are interested in the average quantity, we cannot assume that each mode looks like a string even if $NP \propto L$~\cite{Shimada2021Spatial}.
}.
Thus, the mobility edge, the border between these two species of modes, is defined as a fixed point of $NP/L \propto NP/N^{1/3}=N^{2/3}P$, by changing $N$~\cite{Shklovskii1993Statistics,Coslovich2019localization}.
Figure~\ref{fig:edge} shows $N^{2/3}P$ as functions of the eigenvalue $\lambda$ and the values of $\log_{10}W$ are (a) $-1.5$, (b) $-2$, (c) $-2.5$, (d) $-3$, (e) $-3.5$, (f) $-4$, (g) $-4.5$, and (h) $-5$.
We will show scatter plots of $PN^{2/3}$ versus $\lambda$ in Appendix~\ref{sec:raw}.
The mobility edge, estimated as the mean intersection of pairs of data for different $N$, is shown by the solid line in each of Fig~\ref{fig:edge}(a--d).
Figure~\ref{fig:edge} shows that the modes are localized for $\lambda < \lambda_e$ and delocalized for $\lambda > \lambda_e$.
It can be seen that the mobility edge converges to zero $\lambda_e \to 0$ long before the system falls into an inherent structure.
See also Fig.~\ref{fig:edge_W}, in which we plot the mobility edge as a function of $W$.
This result suggests that all unstable modes are localized during most of the relaxation process in the limit $N \to \infty$; namely, localized modes dominate the dynamics.
Note that the mobility edge cannot be positive because of the phonons that are always delocalized.
To further corroborate this result, it is helpful to compute the level spacing distribution~\cite{Ciliberti2004Localization,Clapa2012Localization,Coslovich2019localization}, although the number of samples used in this study is too small to obtain precise data, particularly for small $W$.

\subsection{The most unstable mode}\label{sec:most}

In Sec.~\ref{sec:spectrum}, we found that the steepest descent dynamics is controlled mostly by unstable localized modes.
Here we focus on the most unstable mode, which is the most localized.
The most unstable mode of a configuration is often well isolated from the other modes~\cite{Lerner2016Statistics,Shimada2018Spatial,Shimada2021Spatial} and thus one can extract characteristics of localized modes cleanly.
We investigate the time evolution of the most unstable mode and how it is stabilized.
Obviously, the most unstable mode is the last unstable mode to disappear and thus its understanding gives us insight into what happens in the late stage of the dynamics.

\subsubsection{Decay profile and correlation length}

\begin{figure}
    \centering
    \includegraphics{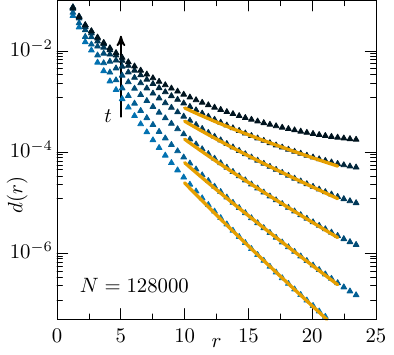}
    \caption{
    Decay profiles for $N=128000$.
    The values of $W$ are $\log_{10}W=-7$, $-6$, $-5$, $-4$, $-3$, and $-2$ from top to bottom.
    The arrow indicates the direction of time.
    The errors are smaller than the symbols and thus not shown.
    The solid lines indicate Eq.~\eqref{Green's function}.
    }
    \label{fig:decay}
\end{figure}

First, we consider the decay profile $d(r)$\footnote{
Strictly speaking, this function needs the mode label $\alpha$ because it is defined for each eigenmode.
However, we focus on the most unstable mode of each configuration and thus the mode label is omitted hereafter.
This also applies to the energy profile.
}~\cite{Gartner2016Nonlinear,Lerner2016Statistics}.
It quantifies how an eigenvector decays from its localized core, defined as the particle with the largest norm (see Appendix~\ref{sec:definitions} for the detailed definition).
Figure~\ref{fig:decay} shows the decay profiles of the most unstable modes for $N=128000$ at several values of $W$.
As in Fig.~\ref{fig:dos_8k}, the arrow shows the direction of time, which is the opposite of $W$.
We present the corresponding data for $N=8000$ and 32000 in Appendix~\ref{sec:system} to examine the system size dependence.
In the early stage of the dynamics, or the large-$W$ regime, the profile decays exponentially with distance $r$ from the core.
This behavior resembles that of unstable localized modes found at saddle points~\cite{Shimada2021Spatial}.
As $W$ decreases with time, this decay becomes slower and the far field develops.

\begin{figure}
    \centering
    \includegraphics{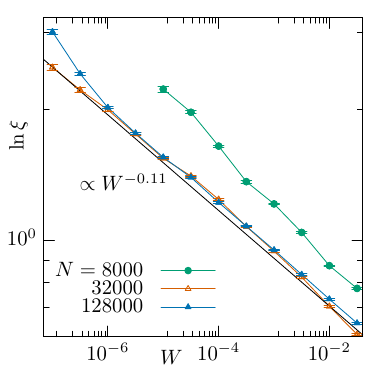}
    \caption{
    The logarithm of the correlation length $\ln \xi$ as a function of the mean squared force $W$.
    The solid line represents a stretched exponential law $\ln \xi \propto W^{-0.11}$.
    }
    \label{fig:decayfit}
\end{figure}

In the case of inherent structures, the lowest-frequency modes are quasi-localized and it has been pointed out that they resemble the elastic response to a local dipolar force~\cite{Lerner2018characteristic,Lerner2021Low,Shimada2020Vibrational}, which is the gradient of Green's function, or the displacement correlation function, of an elastic medium~\cite{Lerner2014Breakdown}.
This fact motivates us to fit the decay profiles in Fig.~\ref{fig:decay} using Green's function.
Since one cannot directly use Green's function of an elastic medium to fit the data of non-equilibrium transient states during the relaxation dynamics, we tentatively assume that Green's function is written in the simplest form $G(r) \sim e^{-r/\xi}/r$, where $\xi$ is the correlation length~\cite{Hansen2013Theory}.
Then the decay profile is expected to follow
\eq{
    d (r) \sim \qty | \pdv{r} G(r) | \sim \qty ( \frac{1}{r^2} + \frac{1}{\xi r} ) e^{-r/\xi} . \label{Green's function}
}
As will be seen below, the correlation length diverges in the long-time limit $\xi \xrightarrow{t\to\infty} \infty$.
This is consistent with the result of inherent structures $d_{\mr{inh}}(r) \sim 1/r^2$, which corresponds to the elastic field of Eshelby inclusion~\cite{Lerner2021Low}.
Thus so to speak, the system behaves like a solid for $r\ll\xi$ and a liquid for $r\gg\xi$.

We fitted Eq.~\eqref{Green's function} to the data in Fig.~\ref{fig:decay}, see the solid lines.
Although the fitting was performed over a range of $r\in[15,20]$, the solid lines are shown over a larger range of $r\in[10,22]$.
We plotted the logarithm of the correlation length $\ln \xi$ as a function of the mean squared force $W$ for $N=8000$, 32000, and 128000 in Fig.~\ref{fig:decayfit}.
The data depend on the system size, but appear to converge rapidly as $N\to\infty$ except for the smallest-$W$ regime, $W\sim 10^{-7}$.
Larger systems are necessary to investigate this regime.
For all $N$, $\xi$ diverges as $W\to0$, or $t\to\infty$, and it obeys a stretched exponential law $\xi \propto e^{W^{-0.11}}$ over four decades in $W$ for $N=128000$.
This simply means that the solid-like region expands during the relaxation process. 
However, this non-trivial time dependence has not been reported in the context of relaxation dynamics, at least to our knowledge, and its underlying mechanism remains to be explored in future work.

\subsubsection{Energy profile and its statistics}

\begin{figure}
    \centering
    \includegraphics{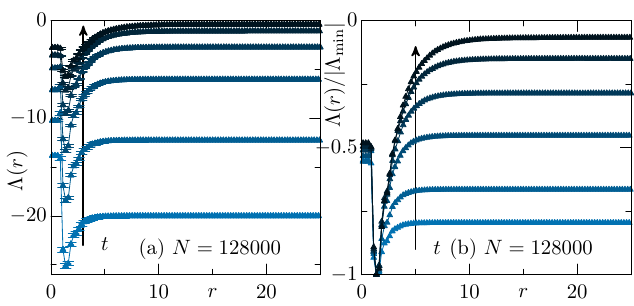}
    \caption{
    (a) Energy profiles for $N=128000$.
    The values of $W$ are $\log_{10}W=-7$, $-6$, $-5$, $-4$, $-3$, and $-2$ from top to bottom.
    The arrow indicates the direction of time.
    (b) Energy profiles normalized by the minimum values $\Lambda_{\mr{min}}$.
    }
    \label{fig:energy}
\end{figure}

It is possible to investigate the local mechanical stability based on the spatial distribution of the eigenenergies of unstable modes.
Thus, we next study the energy profile $\Lambda(r)$~\cite{Shimada2018Spatial,Shimada2021Spatial,Shimada2020Gas} of the most unstable mode.
It is a function of the distance $r$ from the particle that has the most negative local harmonic energy $\delta E_i$ for a given mode~\cite{Shimada2018Spatial}\footnote{In normal mode analysis, one focuses on the harmonic energy. Thus the positive (negative) energy means that the system is (un)stable against small perturbations.}.
Note that the total sum of the local energies is equal to the eigenenergy $\lambda = \sum_i \delta E_{i}$.
The particle with the most negative local energy usually has the largest norm and thus it can also be understood as a core.
Also, the data in Appendix~\ref{sec:number} suggest that the most unstable mode of each configuration has only a single core in terms of the local energy.
The energy profile $\Lambda(r)$ is the partial harmonic energy of a mode integrated from the core to the distance $r$.
In other words, $\Lambda(r)$ is the harmonic energy that the mode would have if the system were cut at a distance $r$ from the core, and $\Lambda(r\to\infty)$ converges to the eigenvalue of the mode (see Appendix~\ref{sec:definitions} for the detailed definition).

Figure~\ref{fig:energy} presents the results for $N=128000$ at the same values of $W$ as in Fig.~\ref{fig:decay}.
Figure~\ref{fig:energy}(a) shows the average energy profile of the most unstable modes.
Roughly speaking, the overall shape of the profile does not change within this range of $W$ and it only shifts upward as a whole with decreasing $W$.
It starts from a negative value by definition, reaches a minimum $\Lambda_{\mr{min}}$, shows an upturn, and finally converges to the average eigenenergy.
The length at which the energy profile reaches its minimum $\Lambda_{\mr{min}}$ is a useful definition of the localization length.
The unstable localized modes are interpreted as defects with the energy scale $\Lambda_{\mr{min}}$, similar to the modes observed in saddle configurations~\cite{Coslovich2019localization,Shimada2021Spatial}.
Figure~\ref{fig:energy}(a) indicates that the localization length of the most unstable mode is determined in the early stage of the dynamics.
This behavior should be compared with the results in Ref.~\cite{Chacko2019Slow}, that is, the size of the hot spots increases with time.
Probably non-affine displacements in a hot spot involve surrounding particles and thus its size appears to grow in the late stage because the surrounding far field develops as shown in Fig.~\ref{fig:decay}.

Also, one can see that the core and far field evolve on different timescales.
Figure~\ref{fig:energy}(b) shows the energy profile normalized by the minimum value $|\Lambda_{\mr{min}}|$.
While the data with different $W$ almost collapse around the origin, they shift upward far from it.
Thus, the far field of the most unstable mode is stabilized more rapidly than the core.

\begin{figure}
    \centering
    \includegraphics{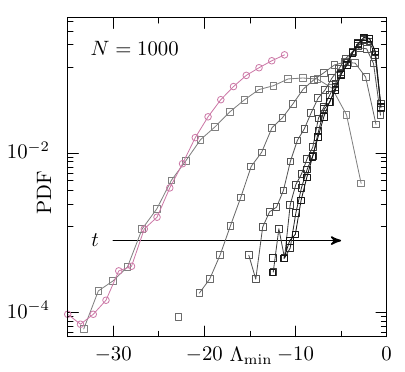}
    \caption{
    PDFs of $\Lambda_{\mr{min}}$ shown by the black squares for $N=1000$.
    The values of $W$ are $\log_{10}W=-9$, $-7.5$, $-6$, $-4.5$, $-3$, and $-1.5$ from right to left.
    The arrow indicates the direction of time.
    The shifted spectrum $100D(\lambda+15.5)$ for $\log_{10}W=-1.5$ is also shown by the pink circles.
    }
    \label{fig:coredist}
\end{figure}

Finally, we analyze the core energy $\Lambda_{\mr{min}}$ in detail.
Figure~\ref{fig:coredist} shows the probability distribution functions (PDFs) of $\Lambda_{\mr{min}}$ for $N=1000$ by the black squares.
For large values of $W$, the function shows a clear exponential tail.
In particular, the shifted spectrum $100D(\lambda+15.5)$ for $\log_{10}W=-1.5$, indicated by the pink circles, almost collapses onto the corresponding data; the functional form of $D(\lambda)$ is almost determined by the statistics of the core energy $\Lambda_{\mr{min}}$ and the far field only shifts the spectrum.
For small values of $W$, the PDF still appears to decay exponentially.
This should be contrasted with the PDF of the eigenvalues of the lowest-frequency quasi-localized modes of inherent structures, which perfectly obeys a Weibull distribution with a power law $g(\omega) \propto \omega^4$~\cite{Lerner2016Statistics,Mizuno2017Continuum,Lerner2021Low}.

\begin{figure}
    \centering
    \includegraphics{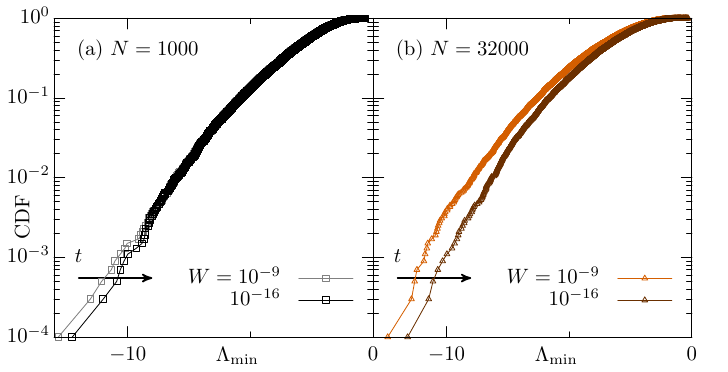}
    \caption{
    CDFs of $\Lambda_{\mr{min}}$ for (a) $N=1000$ and (b) 32000.
    The arrows indicate the direction of time. 
    }
    \label{fig:corecumulative}
\end{figure}

To precisely determine the PDF of the core energy $\Lambda_{\mr{min}}$ for small $W$, we computed its cumulative distribution function (CDF), which is free from the binning error unlike the PDF.
Figure~\ref{fig:corecumulative} shows the CDFs of $\Lambda_{\mr{min}}$ for (a) $N=1000$ and (b) $N=32000$.
The values of $W$ are $10^{-9}$ and $10^{-16}$.
These values of $W$ are very small and the system of $W=10^{-16}$ has no unstable modes, i.e., it is an inherent structure\footnote{
Three out of 10000 configurations of $N=32000$ were stuck at saddles of $W\sim10^{-15}$--$10^{-12}$.
Because each saddle configuration had a single unstable mode, we slightly pushed the configuration along the mode and then the system reached $W=10^{-16}$.
}.
For any $N$ and $W$, the CDF exhibits an exponential tail.
Figures~\ref{fig:coredist} and \ref{fig:corecumulative} show that the distribution of $\Lambda_{\mr{min}}$ remains exponential, whereas that of the eigenenergy $\lambda = \Lambda(r\to\infty)$ gradually develops an algebraic tail~\cite{Lerner2016Statistics,Mizuno2017Continuum,Lerner2021Low}.
For small $W$, the far field essentially contributes to the statistics of the localized modes while it only shifts the spectrum in the large-$W$ regime.

\section{Summary and conclusion}\label{sec:conclusion}

In this paper, we numerically studied the steepest descent dynamics as one of the simplest models of quenching.
We focused on unstable normal modes of instantaneous configurations during the dynamics and investigated how their statistics and spatial structure change toward mechanically stable final configurations called inherent structures.

In Sec.~\ref{sec:mean}, we examined the time evolution of the mean squared force $W$ and the fraction of unstable modes $f_u$.
We compared our data with power laws $W \propto t^{-1.7}$~\cite{Chacko2019Slow,Nishikawa2021Relaxation} and $W \propto f_u^2$ at a sufficiently high temperature $T=2.0$.
These data lead to the vanishing $f_u$-derivative of the total potential energy, $\pdv*{(V/N)}{f_u}\xrightarrow{t\to\infty}0$, meaning that paths in the potential energy landscape passed by quenched liquids become flatter as the dynamics proceeds, i.e., the system moves on a gorge~\cite{Kurchan1996Phase}.
The same analysis of low-temperature configurations revealed that the value of the limit $\lim_{t\to\infty}\pdv*{(V/N)}{f_u}$ does not change.

The results in Sec.~\ref{sec:mean} are intimately related to the notion of marginal stability~\cite{Mueller2015Marginal, Charbonneau2014Fractal,Parisi2020Theory}, which assumes an increasingly important role in understanding anomalous behavior of glasses.
The marginal stability of amorphous materials is often characterized by the flat free energy landscape predicted by replica theory as full replica symmetry breaking~\cite{Charbonneau2014Fractal,Parisi2020Theory}.
In terms of mechanical stability, marginally stable systems have low-energy excitations with a gapless spectrum and are extremely susceptible to perturbations~\cite{Mueller2015Marginal}.
Thus the observed flatness of the potential energy landscape strongly calls for the investigation of other related properties of marginal stability during the relaxation dynamics after quenching.

In Sec.~\ref{sec:spectrum}, we found that the spectrum of unstable normal modes exhibits an exponential tail, a known characteristic of localized modes at saddle points~\cite{Coslovich2019localization}.
Performing a finite-size scaling analysis, we located the mobility edge, which appears to converge to zero long before the system falls into an inherent structure.
Thus, the motion of the particles is mainly governed by localized modes in the late stage of the dynamics.
This finding is consistent with the previous studies~\cite{Chacko2019Slow, Nishikawa2021Relaxation}.

In Sec~\ref{sec:most}, we focused on the most unstable modes of instantaneous configurations and investigated how their structure develops during the relaxation process.
The decay and energy profiles revealed that the core length of these modes is already determined in the early stage of the dynamics.
In contrast, the surrounding far field significantly develops during the dynamics with the correlation length $\xi$, which separates the solid-like and liquid-like regimes.
The correlation length diverges with a non-trivial stretched exponential law $\ln \xi \propto W^{0.11}$.
Moreover, the PDF of the core energy $\Lambda_{\mr{min}}$ remains exponential in contrast to the power-law distribution of the smallest eigenvalues of inherent structures~\cite{Lerner2016Statistics,Mizuno2017Continuum,Lerner2021Low}.
These results mean that the cores of the quasi-localized modes are hot spots left to be unstable during the dynamics while the overall system is rapidly stabilized.
The results in Sec.~\ref{sec:spectrum} and \ref{sec:most} indicate that standard methods of normal mode analysis are quite effective to investigate hot spots, which were originally defined by directly observing particle displacements~\cite{Chacko2019Slow,Nishikawa2021Relaxation}.

To the best of our knowledge, many of the results reported in this paper, such as the stretched exponential law of the correlation length $\xi$, have not been observed in previous studies of the relaxation dynamics and no theory predicts them.
It would be interesting to study whether several theoretical models such as the elastoplastic model discussed in Refs.~\cite{Lin2014density,Lin2014Scaling,Lin2015Criticality,Lin2016Mean} can reproduce them.
Also, it would be worth investigating different quenching algorithms such as the conjugate gradient method or fast inertial relaxation engine~\cite{Bitzek2006Structural}.
Although most studies~\cite{Chacko2019Slow,Nishikawa2021Relaxation}, including this work, have focused on the steepest descent dynamics, which is the overdamped Langevin equation, realistic systems often have inertia and their relaxation dynamics can be modeled as those other algorithms.
Since different algorithms lead to different inherent structures even if the system starts from the same initial state~\cite{Nishikawa2021Relaxation}, it might be the case that some of the results in this paper are modified.
It is then possible to classify the non-equilibrium relaxation dynamics after quenching.

\section*{Conflicts of interest}

There are no conflicts of interest to declare.

\section*{Acknowledgments}

We thank P. Chaudhuri, D. Coslovich, K. Hukushima, Y. Nishikawa, and N. Oyama for useful exchanges.
This work was supported by JSPS KAKENHI Grant Numbers 18H05225, 19H01812, 19J20036, 20H00128, 20H01868, and 21J10021.
Numerical simulations were performed using the Fujitsu PRIMERGY CX400M1/CX2550M5 (Oakbridge-CX) in the Information Technology Center, The University of Tokyo.
This research is partially supported by Initiative on Promotion of Supercomputing for Young or Women Researchers, Information Technology Center, The University of Tokyo.
MS and KS are supported by the JSPS Overseas Research Fellowships.

\clearpage

\appendix

\section{System size dependence}\label{sec:system}

\begin{figure}[h]
    \centering
    \includegraphics{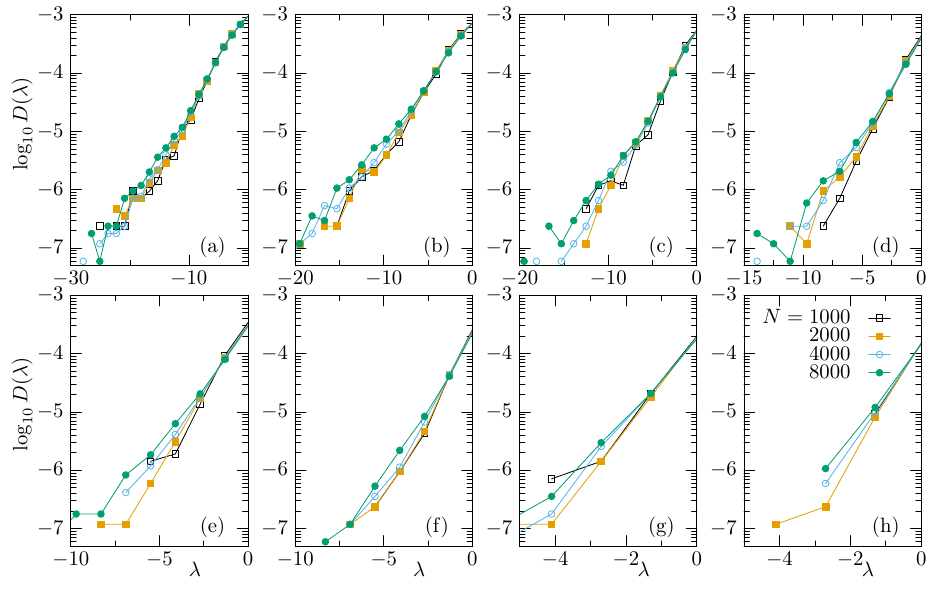}
    \caption{
    Unstable parts of the spectra for $N=1000$, 2000, 4000, and 8000.
    The values of $\log_{10}W$ are (a) $-1.5$, (b) $-2$, (c) $-2.5$, (d) $-3$, (e) $-3.5$, (f) $-4$, (g) $-4.5$, and (h) $-5$.
    }
    \label{fig:dos}
\end{figure}

\begin{figure}
    \centering
    \includegraphics{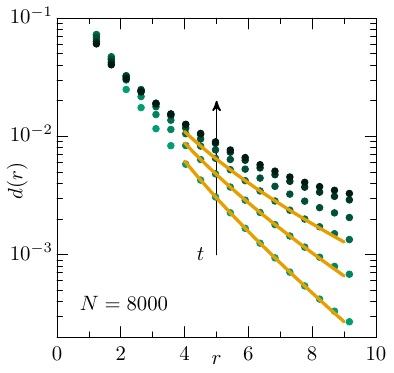}
    \caption{
    Decay profiles for $N=8000$.
    The values of $W$ are $\log_{10}W=-7$, $-6$, $-5$, $-4$, $-3$, and $-2$ from top to bottom.
    The arrow indicates the direction of time.
    The errors are smaller than the symbols and thus not shown.
    The solid lines indicate Eq.~\eqref{Green's function}.
    The fitting was performed over a range of $r\in[5,8]$, but the solid lines are shown over a larger range of $r\in[4,9]$.
    }
    \label{fig:decay_8k}
\end{figure}

\begin{figure}
    \centering
    \includegraphics{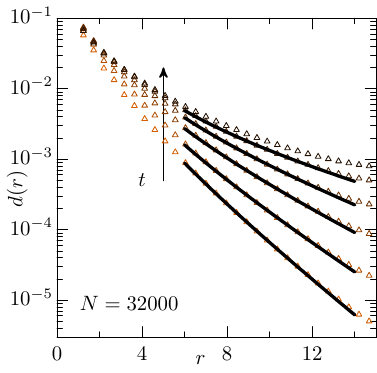}
    \caption{    
    Decay profiles for $N=32000$.
    The values of $W$ are $\log_{10}W=-7$, $-6$, $-5$, $-4$, $-3$, and $-2$ from top to bottom.
    The arrow indicates the direction of time.
    The errors are smaller than the symbols and thus not shown.
    The solid lines indicate Eq.~\eqref{Green's function}.
    The fitting was performed over a range of $r\in[8,12]$, but the solid lines are shown over a larger range of $r\in[6,14]$.
    }
    \label{fig:decay_32k}
\end{figure}

\begin{figure}
    \centering
    \includegraphics{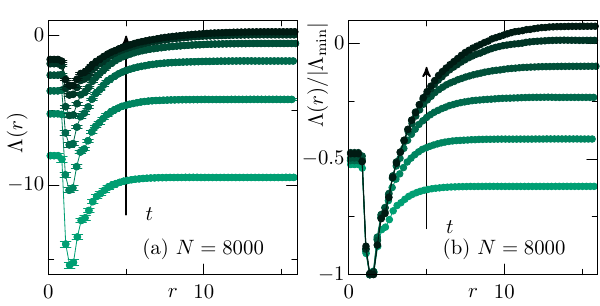}
    \caption{
    (a) Energy profiles for $N=8000$.
    The values of $W$ are $\log_{10}W=-7$, $-6$, $-5$, $-4$, $-3$, and $-2$ from top to bottom.
    The arrow indicates the direction of time.
    (b) Energy profiles normalized by the minimum values $\Lambda_{\mr{min}}$.
    }
    \label{fig:energy_8k}
\end{figure}

\begin{figure}
    \centering
    \includegraphics{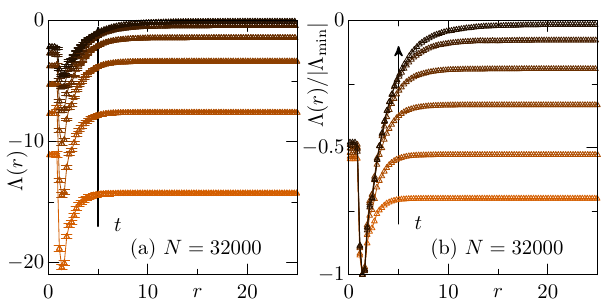}
    \caption{
    (a) Energy profiles for $N=32000$.
    The values of $W$ are $\log_{10}W=-7$, $-6$, $-5$, $-4$, $-3$, and $-2$ from top to bottom.
    The arrow indicates the direction of time.
    (b) Energy profiles normalized by the minimum values $\Lambda_{\mr{min}}$.
    }
    \label{fig:energy_32k}
\end{figure}

We examine the system size dependence of the spectrum, decay profile, and energy profile.
Figure~\ref{fig:dos} shows the spectra for $N=1000$, 2000, 4000, and 8000 at several values of $W$.
In Figs.~\ref{fig:decay_8k} and \ref{fig:decay_32k}, we show the decay profiles for $N=8000$ and 32000 corresponding to Fig.~\ref{fig:decay}.
In Figs.~\ref{fig:energy_8k} and \ref{fig:energy_32k}, we show the energy profiles for $N=8000$ and 32000 corresponding to Fig.~\ref{fig:energy}.
From these figures, it is evident that the data do not qualitatively depend on the system size.

\clearpage

\section{Raw data for the participation ratio}\label{sec:raw}

\begin{figure}[h]
    \centering
    \includegraphics{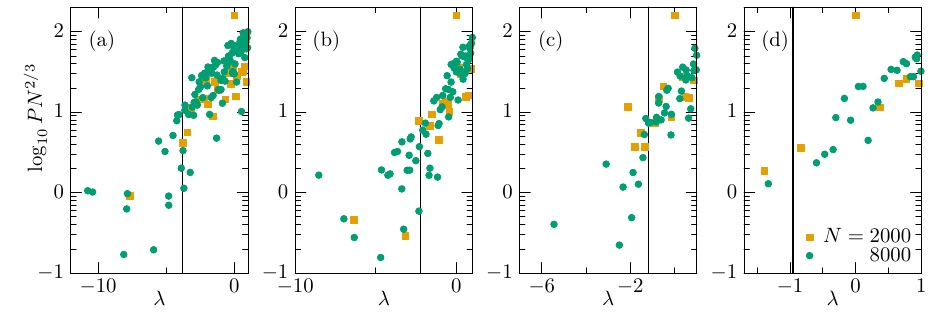}
    \caption{
    Raw data corresponding to Fig.~\ref{fig:edge}.
    The solid lines indicate the mobility edges defined in Sec.~\ref{sec:spectrum}.
    }
    \label{fig:pr_raw}
\end{figure}

Figure~\ref{fig:pr_raw} shows raw data corresponding to Fig.~\ref{fig:edge}(a--d).
We show the data of a single sample for each $N$ and $W$.
The data points look quite scattered, but one can obtain clean data by averaging over a large number of samples as in Fig.~\ref{fig:edge}

\clearpage

\section{Definitions of decay and energy profiles}\label{sec:definitions}

We give the detailed definitions of the decay and energy profiles introduced in Sec.~\ref{sec:most}.

\subsection{Decay profile}

The decay profile~\cite{Lerner2016Statistics,Gartner2016Nonlinear} $d(r)$\footnote{
Note that the mode label $\alpha$ is omitted here as in Sec.~\ref{sec:most}.}
of a mode $\bs{e}$ is a function of the distance $r$ from the particle with the largest norm $|\bs{e}_i|$, the position of which is denoted by $\bs{r}_d$.
It is defined as
\eq{
    d(r) = \underset{\substack{i \\ |\bs{r}_i - \bs{r}_d|\in[r-\Delta r/2,r+\Delta r/2]}}{\operatorname{median}}|\bs{e}_i| ,
}
where $\Delta r$ is the bin width.
The distance $r$ used here differs slightly from the one in the original paper by Gartner and Lerner~\cite{Gartner2016Nonlinear}. 
They defined the center of an eigenmode using $w$ particles with the largest norms and measured the distance $r$ from this center.
They mainly used $w=4$ while our definition corresponds to $w=1$.
However, this difference does not affect the results qualitatively in the case of spatially localized modes.

\subsection{Energy profile}

The local vibrational energy of the $i$-th particle in a mode $\bs{e}$ is
\eq{
    \delta E_i & = \frac{1}{2}\sum_{j=1}^N\left[v''_{ij}(r_{ij})(\bs{n}_{ij}\cdot\bs{e}_{ij})^2 + \frac{v'_{ij}(r_{ij})}{r_{ij}}(\bs{e}_{ij}^\perp)^2\right] , 
}
where $\bs{n}_{ij}$ is the unit vector pointing from $i$ to $j$, $\bs{e}_{ij}=\bs{e}_i-\bs{e}_j$ is the relative displacement between the pair, and $(\bs{e}_{ij}^\perp)^2 = (\bs{e}_{ij})^2-(\bs{n}_{ij}\cdot\bs{e}_{ij})^2$ is the squared transverse relative displacement. 
The energy profile~\cite{Shimada2018Spatial} $\Lambda(r)$ is a function of the distance $r$ from the particle with the most negative $\delta E_i$, the position of which is denoted by $\bs{r}_e$.
It is defined as
\eq{
    \Lambda(r) & = \sum_{i = 1}^N \Theta(r-|\bs{r}_i-\bs{r}_e|) \delta E_i = \int_{|\bs{x}| < r} d\bs{x} \delta E (\bs{x}), 
}
where $\Theta(x)$ is the Heaviside step function.
In the rightmost expression, we rewrote the function using a spatial integral of the local energy density $\delta E(\bs{r}) = \sum_i \delta E_i \delta[ \bs{r} - (\bs{r}_i-\bs{r}_e) ]$.
Thus, the energy profile is the vibrational energy that the mode would have if the system was cut at a distance $r$ from the center $\bs{r}_e$.
It converges to the eigenvalue of the mode $\lambda$ as $r\to\infty$.
Finally, note that the difference between $\bs{r}_d$ and $\bs{r}_e$ does not matter in practice because they usually coincide~\cite{Shimada2018Spatial}.

\clearpage

\section{Number of cores}\label{sec:number}

\begin{figure}[ht]
    \centering
    \includegraphics{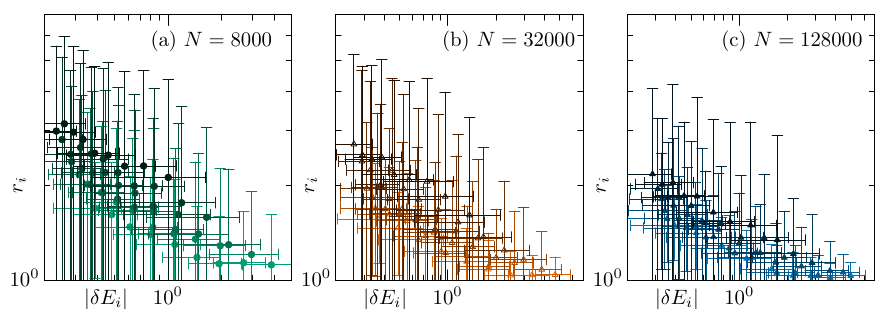}
    \caption{
    The average distance from the core $r_i$ versus the harmonic energy $\delta E_i$ for (a) $N=8000$, (b) 32000, and (c) 128000.
    The error bar indicates the standard deviation of each data point and not the error itself.
    The values of $W$ are $\log_{10}W = -2$, $-3$, $-4$, $-5$, $-6$, and $-7$ from light to dark colors.
    }
    \label{fig:coredistance}
\end{figure}

We show data that suggest that there is a single core in the most unstable mode of each configuration on average.
By the term ``core'', we mean the particle with the most negative harmonic energy $\delta E_i$.
In this appendix, the particles are sorted in the ascending order of $\delta E_i$, i.e., $\delta E_1 < \delta E_2 < \cdots$.
In Fig.~\ref{fig:coredistance}, the average distance from the core $r_i$ is plotted against the average harmonic energy $\delta E_i$ for several values of $W$ indicated by the colors.
Each panel shows only the data for the ten particles with the most negative harmonic energies for each $W$.
This figure indicates that these unstable particles are close to the core particles $i=1$ and thus, it is unlikely that they constitute other independent cores.

\clearpage

\section{Weakly jammed packings}\label{sec:jammed}

We investigate instantaneous normal modes of weakly jammed packings during the relaxation dynamics after quenching for comparison with the results of the KALJ model in the main text.
This system is composed of monodisperse athermal particles with mass $m$ interacting via a purely repulsive potential.
Here we adopted a harmonic potential 
\eq{
    v(r) = \frac{\epsilon}{2} \qty ( 1 - \frac{r}{\sigma} )^2 \Theta \qty ( 1 - \frac{r}{\sigma} ) .
}
In this section, length, mass, and time are reported in units of $\sigma$, $m$, and $\sqrt{m \sigma^2/\epsilon}$, respectively.
The packing fraction is fixed at the same value $\phi=0.73$ as in Ref.~\cite{Nishikawa2021Relaxation}.
Starting from random configurations, which are equivalent to infinite-temperature states, we performed simulations of the steepest descent dynamics in Eq.~\eqref{steepest descent}.
We repeated some of the analysis in the main text.
The number of configurations used for each $N$ is listed in Tab.~\ref{tab:hs_samples}.

\begin{table}[h]
    \renewcommand{\arraystretch}{1.2}
    \setlength\tabcolsep{0.5em}
    \centering
    \caption{
    The number of samples for each $N$.
    }
    \begin{tabular}{cc} \hline \hline
        $N$     & \#samples     \\ \hline
        2000    & 1000          \\
        8000    & 500           \\
        32000   & 200           \\
        128000  & 100           \\ \hline \hline
    \end{tabular}
    \label{tab:hs_samples}
\end{table}

\subsection{Mean squared force and fraction of unstable modes}

\begin{figure}[h]
    \centering
    \includegraphics{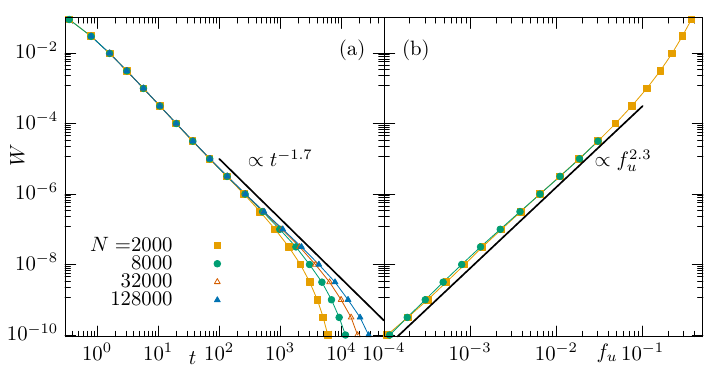}
    \caption{
    (a) Average mean squared force as functions of the time.
    The solid line represents a power law $W \propto t^{-1.7}$~\cite{Chacko2019Slow,Nishikawa2021Relaxation}.
    Note that $W$ is controlled in this plot.
    See the text for details.
    The errors are smaller than the symbols and thus not shown.
    (b) Average mean squared force versus fraction of unstable modes.
    The solid line represents a power law $W \propto f_u^{2.3}$.
    }
    \label{fig:hs_force}
\end{figure}

Figure~\ref{fig:hs_force} corresponds to Fig.~\ref{fig:force}.
One can observe power laws $W \propto t^{-1.7}$ and $W \propto f_u^{2.3}$.
The former is already known~\cite{Chacko2019Slow,Nishikawa2021Relaxation}.
Although the exponent of the latter power law is different from those of the KALJ model, it is easy to repeat the same discussion in Sec.~\ref{sec:mean}, which leads to 
\eq{
    \pdv{(V/N)}{f_u} \xrightarrow{t\to\infty} 0 .
}

\subsection{Spectrum and localization}

\begin{figure}[h]
    \centering
    \includegraphics{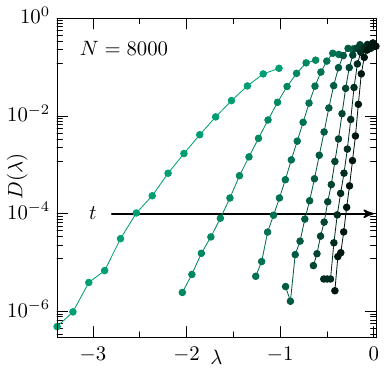}
    \caption{
    Unstable parts of the spectra for $N=8000$ during the dynamics.
    The values of the mean squared force are $\log_{10}W = -4.5,-4.0,-3.5,-3.0,-2.5,-2.0$, and $-1.5$ from right to left.
    The time is indicated by the arrow, which is the opposite of $W$.
    }
    \label{fig:hs_dos_8k}
\end{figure}

\begin{figure}[h]
    \centering
    \includegraphics{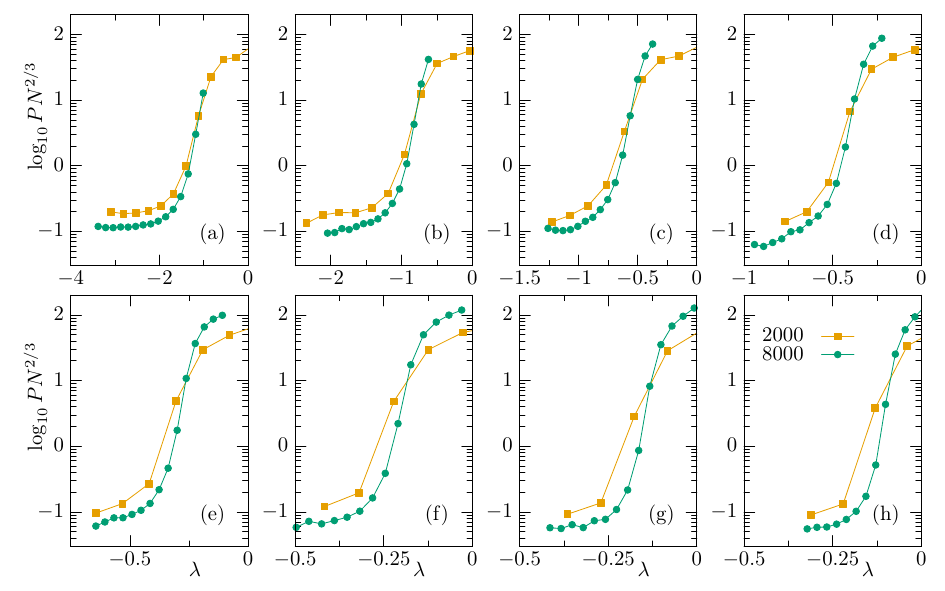}
    \caption{
    $\log_{10}PN^{2/3}$ versus $\lambda$.
    The values of $\log_{10}W$ are (a) $-1.5$, (b) $-2$, (c) $-2.5$, (d) $-3$, (e) $-3.5$, (f) $-4$, (g) $-4.5$, and (h) $-5$.
    }
    \label{fig:hs_edge}
\end{figure}

Figures~\ref{fig:hs_dos_8k} and \ref{fig:hs_edge} correspond to Figs.~\ref{fig:dos_8k} and \ref{fig:edge}, respectively.
Although we did not compute the mobility edge, it is obvious from Fig.~\ref{fig:hs_edge} that the delocalized modes gradually disappear.

\subsection{The most unstable mode}

\begin{figure}[h]
    \centering
    \includegraphics{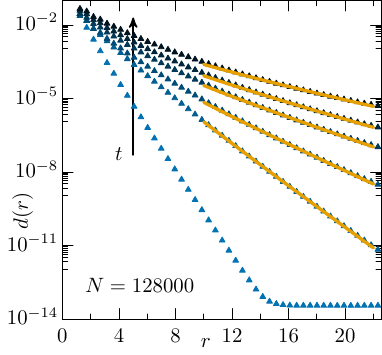}
    \caption{
    Decay profiles for $N=128000$.
    The values of $W$ are $\log_{10}W=-10$, $-9$, $-8$, $-6$, $-4$, and $-2$ from top to bottom.
    The arrow indicates the direction of time.
    The errors are smaller than the symbols and thus not shown.
    The solid lines indicate Eq.~\eqref{Green's function}.
    The fitting was performed over a range of $r\in[15,20]$, but the solid lines are shown over a larger range of $r\in[10,22]$.
    }
    \label{fig:hs_decay}
\end{figure}

\begin{figure}[h]
    \centering
    \includegraphics{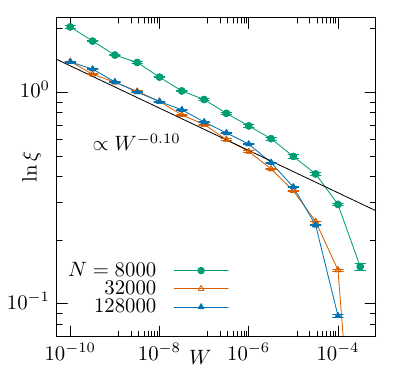}
    \caption{
    The logarithm of the correlation length $\ln \xi$ as a function of the mean squared force $W$.
    The solid line represents a stretched exponential law $\ln \xi \propto W^{-0.10}$.
    }
    \label{fig:hs_decayfit}
\end{figure}

\begin{figure}[ht]
    \centering
    \includegraphics{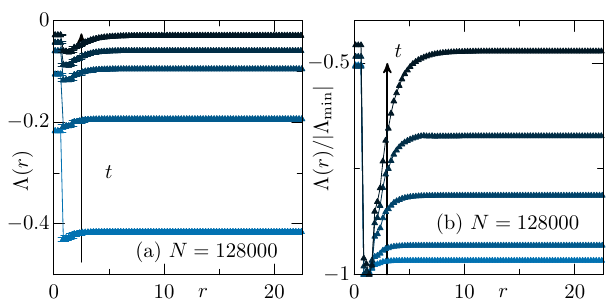}
    \caption{
    (a) Energy profiles for $N=128000$.
    The values of $W$ are $\log_{10}W=-10$, $-9$, $-8$, $-6$, and $-4$ from top to bottom.
    The arrow indicates the direction of time.
    (b) Energy profiles normalized by the minimum values $\Lambda_{\mr{min}}$.
    }
    \label{fig:hs_energy}
\end{figure}

Figures~\ref{fig:hs_decay}, \ref{fig:hs_decayfit}, and \ref{fig:hs_energy} correspond to Figs.~\ref{fig:decay}, \ref{fig:decayfit}, and \ref{fig:energy}, respectively.
In general, the behavior of these functions is qualitatively the same as that for the KALJ model, but the decay profile decays more rapidly in the early stage of the dynamics, or the large-$W$ regime.
As a result, the correlation length becomes shorter and does not obey the stretched exponential law $\ln\xi \propto W^{0.10}$ for large $W$.
A possible explanation is that a contact network of particles does not percolate in this early stage; namely, the system is composed of small clusters of particles.
Such states are not possible in systems with long-range interactions including the KALJ model.

\newpage

\subsection{System size dependence}

\begin{figure}[h]
    \centering
    \includegraphics{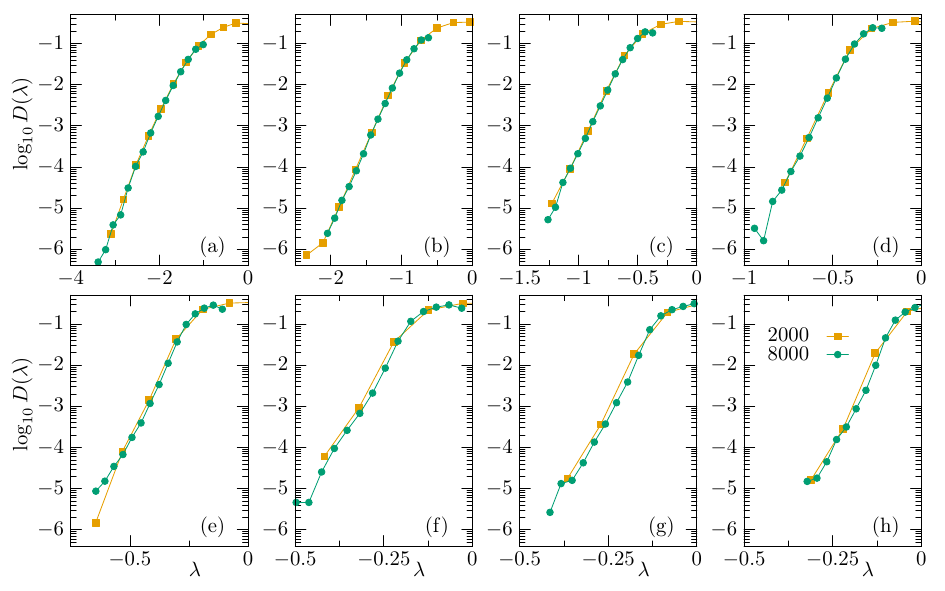}
    \caption{
    Unstable parts of the spectra for $N=2000$ and 8000.
    The values of $\log_{10}W$ are (a) $-1.5$, (b) $-2$, (c) $-2.5$, (d) $-3$, (e) $-3.5$, (f) $-4$, (g) $-4.5$, and (h) $-5$.
    }
    \label{fig:hs_dos}
\end{figure}

\begin{figure}
    \centering
    \includegraphics{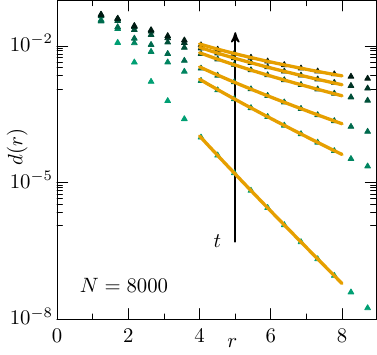}
    \caption{
    Decay profiles for $N=8000$.
    The values of $W$ are $\log_{10}W=-10$, $-9$, $-8$, $-6$, $-4$, and $-2$ from top to bottom.
    The arrow indicates the direction of time.
    The errors are smaller than the symbols and thus not shown.
    The solid lines indicate Eq.~\eqref{Green's function}.
    The fitting was performed over a range of $r\in[5,7.5]$, but the solid lines are shown over a larger range of $r\in[4,8]$.
    }
    \label{fig:hs_decay_8k}
\end{figure}

\begin{figure}
    \centering
    \includegraphics{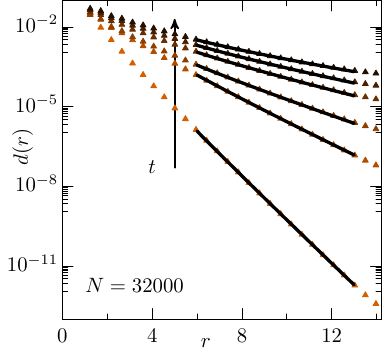}
    \caption{
    Decay profiles for $N=32000$.
    The values of $W$ are $\log_{10}W=-10$, $-9$, $-8$, $-6$, $-4$, and $-2$ from top to bottom.
    The arrow indicates the direction of time.
    The errors are smaller than the symbols and thus not shown.
    The solid lines indicate Eq.~\eqref{Green's function}.
    The fitting was performed over a range of $r\in[8,12]$, but the solid lines are shown over a larger range of $r\in[6,13]$.
    }
    \label{fig:hs_decay_32k}
\end{figure}

\begin{figure}
    \centering
    \includegraphics{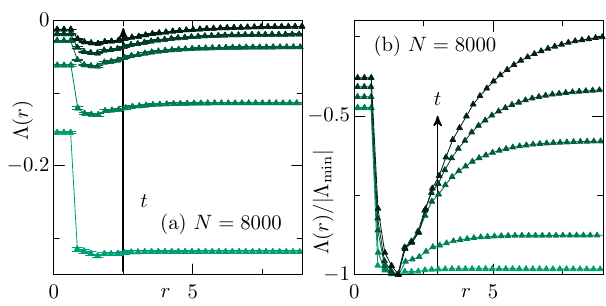}
    \caption{
    (a) Energy profiles for $N=8000$.
    The values of $W$ are $\log_{10}W=-10$, $-9$, $-8$, $-6$, and $-4$ from top to bottom.
    The arrow indicates the direction of time.
    (b) Energy profiles normalized by the minimum values $\Lambda_{\mr{min}}$.
    }
    \label{fig:hs_energy_8k}
\end{figure}

\begin{figure}
    \centering
    \includegraphics{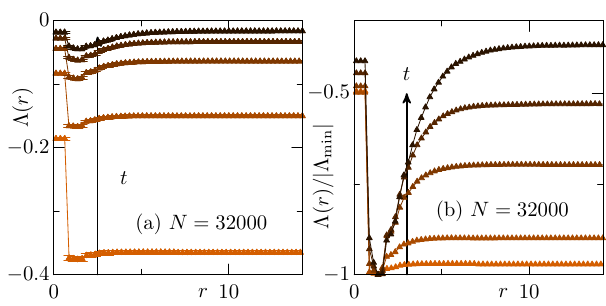}
    \caption{
    (a) Energy profiles for $N=32000$.
    The values of $W$ are $\log_{10}W=-10$, $-9$, $-8$, $-6$, and $-4$ from top to bottom.
    The arrow indicates the direction of time.
    (b) Energy profiles normalized by the minimum values $\Lambda_{\mr{min}}$.
    }
    \label{fig:hs_energy_32k}
\end{figure}

Figures~\ref{fig:hs_dos}, \ref{fig:hs_decay_8k}, \ref{fig:hs_decay_32k}, \ref{fig:hs_energy_8k}, and \ref{fig:hs_energy_32k} correspond to Figs.~\ref{fig:dos}, \ref{fig:decay_8k}, \ref{fig:decay_32k}, \ref{fig:energy_8k}, and \ref{fig:energy_32k}, respectively.
One can see that the data do not depend on the system size qualitatively.

\clearpage

\end{document}